\documentclass[11pt,a4paper]{article}
\usepackage{amsmath}
\usepackage{amssymb}
\usepackage[tight]{units}
\usepackage{textcomp}
\usepackage{array}
\usepackage[numbers]{natbib}
\usepackage{graphicx}
\usepackage{subfig}
\usepackage[affil-it]{authblk}

\begin{document} 

\title{\LARGE \bf Probing low WIMP masses with the next generation of CRESST detectors}

\author[a]{G.~Angloher}
\author[a,b]{A.~Bento}
\author[c]{C.~Bucci}
\author[c]{L.~Canonica}
\author[d,e]{A.~Erb}
\author[d]{F.~v.~Feilitzsch}
\author[a]{N.~Ferreiro~Iachellini}
\author[c]{P.~Gorla}
\author[f]{A.~G\"utlein}
\author[a]{D.~Hauff}
\author[g]{J.~Jochum}
\author[a]{M.~Kiefer}
\author[f]{H.~Kluck}
\author[h]{H.~Kraus}
\author[d]{J.-C.~Lanfranchi}
\author[g]{J.~Loebell}
\author[d]{A.~M\"unster}
\author[a]{F.~Petricca\thanks{corresponding author: petricca@mpp.mpg.de}}
\author[d]{W.~Potzel}
\author[a]{F.~Pr\"obst}
\author[a]{F.~Reindl}
\author[c]{K.~Sch\"affner}
\author[f]{J.~Schieck}
\author[g]{S.~Scholl}
\author[d]{S.~Sch\"onert}
\author[a]{W.~Seidel}
\author[a]{L.~Stodolsky}
\author[g]{C.~Strandhagen}
\author[a]{R.~Strauss}
\author[a]{A.~Tanzke}
\author[g]{M.~Uffinger}
\author[d]{A.~Ulrich}
\author[g]{I.~Usherov}
\author[d]{S.~Wawoczny}
\author[d]{M.~Willers}
\author[a]{M.~W\"ustrich}
\author[d]{A.~Z\"oller}

\affil[a]{Max-Planck-Institut f\"ur Physik, D-80805 M\"unchen, Germany}
\affil[b]{Departamento de Fisica, Universidade de Coimbra, P3004 516 Coimbra, Portugal}
\affil[c]{INFN, Laboratori Nazionali del Gran Sasso, I-67010 Assergi, Italy}
\affil[d]{Physik-Department, Technische Universit\"at M\"unchen, D-85747 Garching, Germany}
\affil[e]{Walther-Mei\ss ner-Institut f\"ur Tieftemperaturforschung, D-85748 Garching, Germany}
\affil[f]{Institut f\"ur Hochenergiephysik der \"Osterreichischen Akademie der Wissenschaften, A-1050 Wien, Austria
and Atominstitut, f University of Technology, A-1020 Wien, Austria}
\affil[g]{Eberhard-Karls-Universit\"at T\"ubingen, D-72076 T\"ubingen, Germany}
\affil[h]{Department of Physics, University of Oxford, Oxford OX1 3RH, United Kingdom}

\renewcommand\Authands{, and }
\renewcommand\Affilfont{\itshape\small}

\maketitle

\begin{abstract}
The purpose of this document is to describe the upgrade of the CRESST dark matter search at LNGS \cite{CRESST_Astropart23,CRESST_Astropart31}. The proposed strategy  will allow to explore a region of the parameter space for spin-independent WIMP-nucleon elastic scattering corresponding to WIMP masses below 10~GeV/c$^\text{2}$, that has not been covered by other experiments. These results can be achieved only  with outstanding detector performances in terms of threshold and background. This proposal shows how CRESST can match these performance requirements, adding a unique piece of information to the dark matter puzzle. The results of this program will fix a new state-of-the-art in the low mass WIMP exploration, opening new perspectives of understanding the dark matter scenario.
\end{abstract}

\newpage

\section{Physics case}
In the last few years many direct dark matter projects have probed the mass-cross section parameter space for WIMP-nucleus elastic scattering\footnote{To compare data from different dark matter direct detection experiments, one needs to assume a model of particle interaction between dark matter particles and nuclei in the detector as well as a model for the dark matter halo. For spin-independent WIMP-nucleus scattering, the cross section $\sigma$ is assumed to be proportional to A$^2$, where A is the mass number of the target nuclei. For the dark matter halo model the standard assumption is an isothermal WIMP halo of density 0.3 GeV/cm$^\text{3}$, with a galactic escape velocity of 544 km/s and an asymptotic velocity of 220 km/s.} with increasing sensitivity \cite{Snowmass}. Most of these experiments are suitable for WIMP masses $\gtrsim$~30~GeV/c$^\text{2}$, where the sensitivity gain is mainly driven by the exposure.\\
Nevertheless, a number of theoretical models favoring lighter WIMP candidates (e.g. \cite{AsymmetricDM_1, AsymmetricDM_2, dipoleDM, dipoleDM_constraints, Isospinviolating}) have recently moved the interest of the community to the mass region $\lesssim$~10~GeV/c$^\text{2}$.\\
As such light WIMPs produce only low-energy nuclear recoils $O$(keV), the challenge for their detection is to achieve a sufficiently low threshold, with enough background discrimination at these low energies.\\
The results described in \cite{CRESST_run33} show that upgraded CRESST-II detectors are, among the different technologies, the most suited for detecting low mass WIMPs. The experiment explored the mass region below 6~GeV/c$^\text{2}$ with a sensitivity that is not accessible to most current generations dark matter experiments.\\
These results can be further improved focusing on the performance of the detectors. We propose a phased program of 3 steps of increasing sensitivity, with the final goal of reaching the coherent neutrino scattering region \cite{Achim_neutrino}.

\section{Current status of the experiment}\label{Current}
\begin{figure}[h!]
 \begin{center}
   \includegraphics[width=0.6\textwidth]{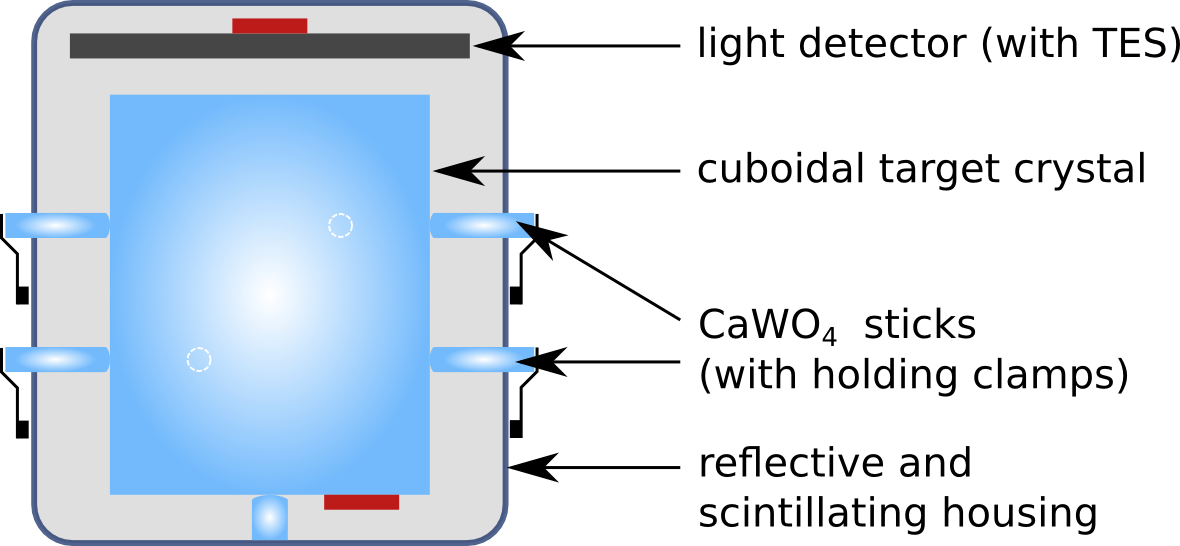}
   \caption{Scheme of the “actively discriminating detector” design which uses CaWO$_\text{4}$ sticks to hold the target crystal.}
   \label{fig:NewDesigns_Stick}
 \end{center}
\end{figure}

The main goal of the current run of the CRESST experiment (run33) was to clarify the nature of the signal excess reported in \cite{CRESST_run32}, reducing neutron, $\alpha$ and  $^\text{206}$Pb recoil backgrounds by more than one order of magnitude.
The experiment has been acquiring data since July 30th 2013. The data set until January 7th 2014 has been used\footnote{This first data set has been analyzed in a non-blind way, namely the data have been used to define quality cuts which are applied to the raw data to ensure that only valid events are considered for further analysis. The data collected afterwards are analyzed in a blind way, i.e. the cuts optimized on the non-blind data are not modified any further.} for a first evaluation of the detector performance and of the impact of the various actions for background reduction which have been implemented for this run.\\ 
The data set shows that the residual neutron background has been efficiently suppressed with the introduction of an additional PE layer inside the Pb/Cu shield. The background of degraded $\alpha$s, which originated from a bulk contamination of the bronze (CuSn6) used for the clamps holding the crystals, has also been completely removed by introducing new holding clamps from purpose-produced, ultra-pure bronze. The $^\text{206}$Pb nuclear recoil background, however, was impossible to suppress with passive techniques such as cleaning and shielding the detectors. This background is efficiently suppressed only in the three new ``actively discriminating detector'' designs that were introduced.  In these modules all the material surrounding the detector is active, allowing to tag $\alpha$ decays from all inner surfaces of the detector module.\\
The data set from a single ``actively discriminating detector'' module, corresponding to an exposure of about 29~kg~days, has been used to derive the low-mass WIMP limit presented in \cite{CRESST_run33}.\\ 
A schematic view of the design of this module is presented in fig.~\ref{fig:NewDesigns_Stick} and a detailed description can be found in \cite{Raimund_TUM40_layout}. The design uses CaWO$_\text{4}$ sticks to hold the crystal. The sticks are held in place by clamps mounted outside the housing. In this way this design has a fully scintillating inner surface which allows to distinguish the background from surface $\alpha$ decays by using the additional light signal of the $\alpha$ as a veto. Further innovation in this design is the use of a cuboidal crystal, instead of a cylindrical one, for improved light collection.

\begin{figure}[h!]
 \begin{center}
   \includegraphics[width=0.65\textwidth]{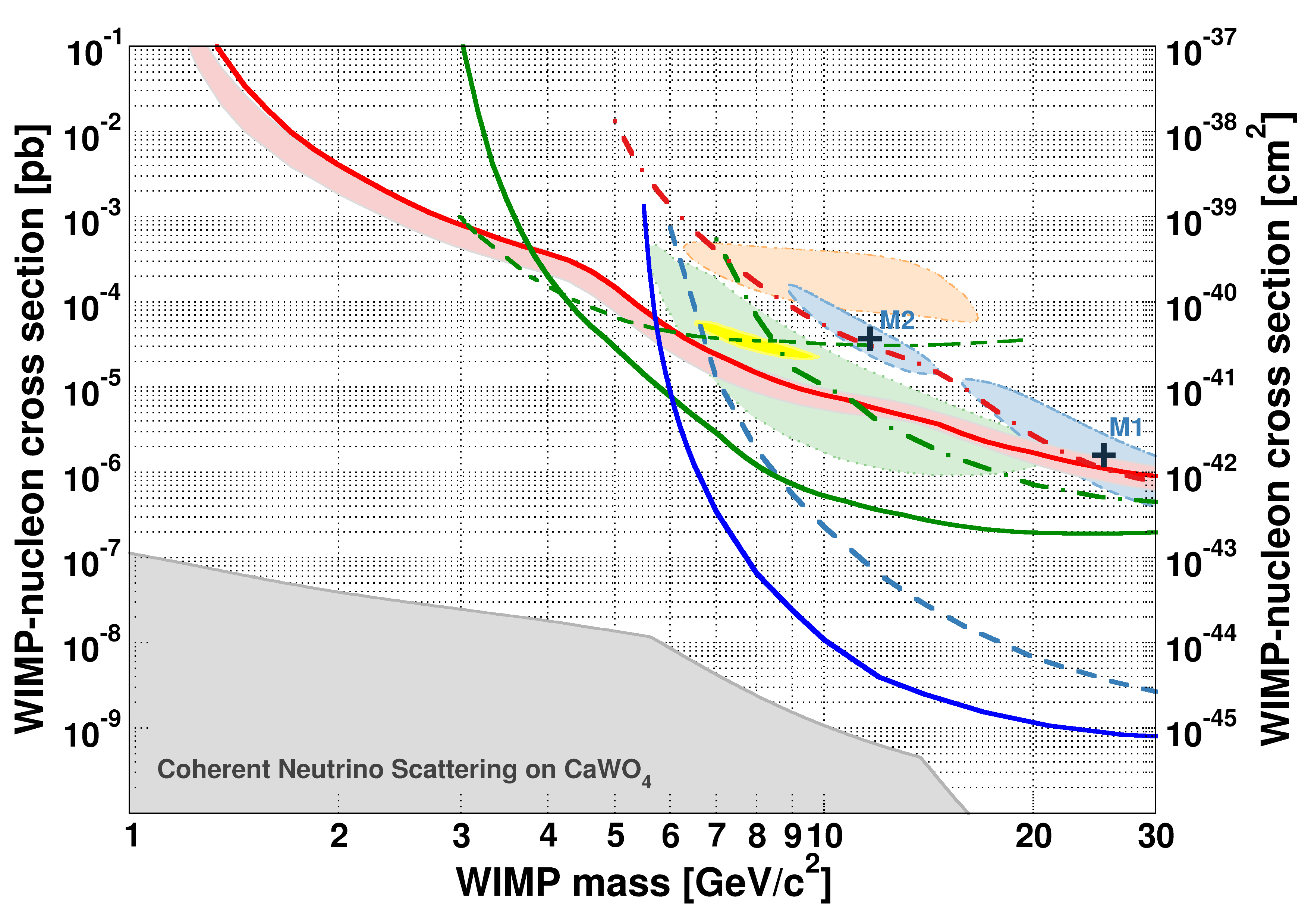}
   \caption{WIMP parameter space for spin-independent ($\sigma \propto$ A$^\text{2}$) WIMP-nucleon scattering. The 90 \% C.L. upper limit (solid red) is depicted together with the expected sensitivity (1 $\sigma$ C.L.) from the background-only model (light red band). The CRESST 2 $\sigma$ contour reported in \cite{CRESST_run33} is shown in light blue. The dash-dotted red line refers to the reanalyzed data from the CRESST commissioning run \cite{Oxford}. Marked in grey is the limit for a background-free CaWO$_\text{4}$ experiment arising from coherent neutrino-nucleus scattering, dominantly from solar neutrinos. Shown in green are the limits (90 \% C.L.) from Ge-based experiments: Super-CDMS (solid) \cite{superCDMS}, CDMSlite (dashed) \cite{CDMSlite} and EDELWEISS (dash-dotted) \cite{EDELWEISS}. The parameter space favored by CDMS-Si \cite{CDMS_Si} is shown in light green (90 \% C.L.), the one favored by CoGeNT (99 \% C.L. \cite{CoGeNT})and DAMA/Libra (3 σ C.L. \cite{DAMA}) in yellow and orange. The exclusion 
curves from liquid xenon experiments (90 \% C.L.) are drawn in blue,solid for LUX \cite{LUX}, dashed for XENON100 \cite{XENON}. Marked in grey is the limit for a background-free CaWO$_\text{4}$ experiment arising from coherent neutrino scattering, dominantly from solar neutrinos \cite{Achim_neutrino}.}
      \label{fig:Exclusion_run33}
 \end{center}
\end{figure}

The CaWO$_\text{4}$ crystal of this module (named TUM40) has been produced at the crystal growing facility of the Technische Universit\"at M\"unchen \cite{TUM_crystals}. It shows a significant gain in terms of radiopurity with respect to the commercially available ones (a factor of 2 to 10 lower $e^-/\gamma$ background in the region of interest with an average rate of about 3.5/[kg keV day] and a significant reduction of the $\alpha$ contamination corresponding to a total intrinsic alpha activity from natural decay chains of about 3 mBq/kg \cite{Andrea, Raimund_TUM40_bck}). This improvement has a great impact on the low energy region, where it allows a significant gain in sensitivity for low WIMP masses which is visible in the limit obtained from the detector presented in fig.~\ref{fig:Exclusion_run33} (compare red solid line with red dash-dotted line).

\begin{figure}[h!]
 \begin{center}
   \includegraphics[width=0.65\textwidth]{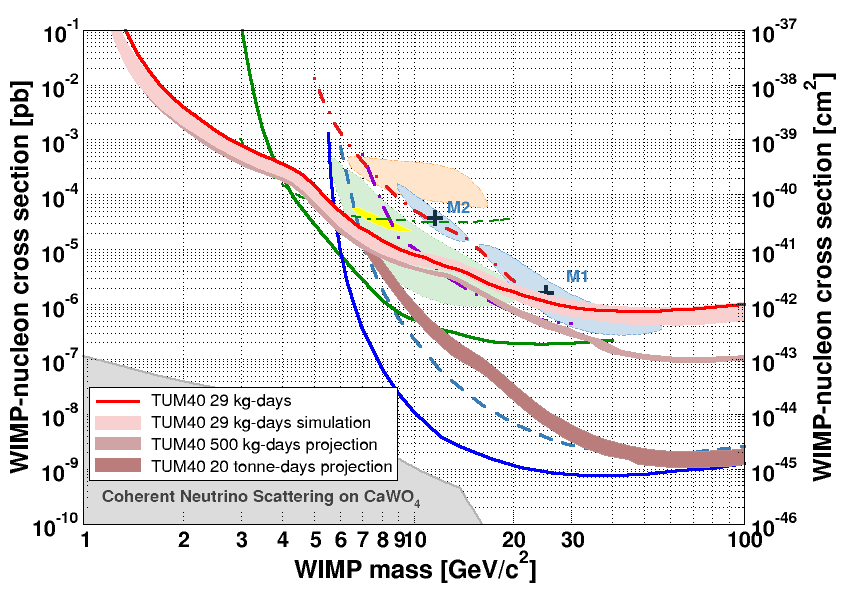}
   \caption{In addition to the limits shown in fig.~\ref{fig:Exclusion_run33}, the expected sensitivity for an exposure of about 500~kg~days and 20~tonne~days with the same detector are shown.}
   \label{fig:Projections_Run33}
 \end{center}
\end{figure}

The CRESST exclusion limit rises more moderately and has lower systematic uncertainty than the limits from other experiments when going to low WIMP masses. These distinctive features, which come from the possibility to simultaneously probe a possible WIMP signal on different nuclei (including the very light oxygen nuclei) and to measure nuclear recoil energies with little systematic uncertainty, mark the uniqueness of the experimental technique \cite{CRESST_run33}.\\
The result from the data obtained already excludes the lower mass maximum (M2), but more statistics is required to improve the limit at higher WIMP masses and, thus, to clarify the origin of the higher mass maximum (M1). For this reason the current run will go on until an exposure of about 500~kg~days is achieved with the actively discriminating modules, which is expected to be achieved in 2015 (see fig.~\ref{fig:Projections_Run33}). Without a gain in performance (e.g. lowering of the threshold), this increased statistics will lead only to a marginal improvement of the result in the low WIMP masses, where the sensitivity is already limited by the detector performance.\\
In the high-mass WIMP region the sensitivity still scales with exposure for state-of-the-art CRESST detectors with the present background level. In fig.~\ref{fig:Projections_Run33} a projection for an exposure of 20 tonne-days\footnote{Such an exposure goal can realistically only be achieved building larger detectors ($O$(1kg)) than the current ones.} is shown. 

\section{Future potential}
Due to the complexity of CRESST detectors, pursuing a gain of sensitivity at high WIMP masses, where large exposures are crucial, requires a significant effort. In recent years liquid-noble-gas detectors showed to be more competitive in terms of mass scalability. On the other hand, the results described in sec.~\ref{Current} demonstrate that the CRESST experiment can be a unique tool to explore low WIMP masses.\\
The sensitivity for light WIMPs can be improved in future runs by enhancing the detector performance, namely lowering the threshold and decreasing the background in the energy range of interest.\\
The background contribution in the energy range of interest can be reduced working on complementary aspects. Considering the case of the result in fig.~\ref{fig:Exclusion_run33}, the limit derived from data and a Monte Carlo simulation which assumes the presence of $e^-/\gamma$-backgrounds only, agree throughout the whole WIMP mass range. This indicates that the events in the acceptance region are likely to be explained as leakage from the $e^-/\gamma$-band. Here, improving the detector performance (i.e. reducing leakage from the $e^-/\gamma$-band) could be obtained by reducing the overall content of the $e^-/\gamma$-band or/and reducing the overlap of the bands.\\
The reduction of the $e^-/\gamma$-band content can be achieved by improving the radiopurity of the crystals (selection of raw materials, multiple growing of crystals \cite{TUM_crystals, Andrea, Raimund_TUM40_bck}) and reducing the background originating from the surrounding of the detectors (e.g. careful selection of materials, upgraded cleaning procedures, minimized exposure to radon, active veto). The overlap of the bands can be reduced acting on a set of parameters: i) amount of detected light (which is a combination of light output of the crystal, light transport in the reflecting housing and light absorbed in the light detectors), ii) non-proportionality of the light yield (decrease of the light yield of the $e^-/\gamma$-band towards lower energies, as observed in most inorganic scintillators\footnote{Although the non-proportionality of the light yield is a physical effect, a detector dependence is observed and there are good indications that improvements are possible for the TUM-grown crystals.}) and 
iii) resolution of the light channel.

\subsection{Optimized detector layout}

\begin{figure}[h!]
 \begin{center}
   \includegraphics[width=0.5\textwidth]{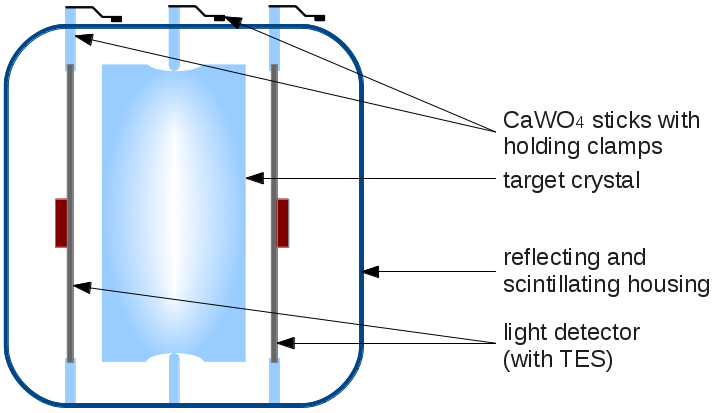}
   \caption{Schematic view of a possible design for a small detector module.}
   \label{fig:Small_module}
 \end{center}
\end{figure}

The current CRESST approach consists in developing an optimized detector layout. This implies a radical change in the strategy, leaving the path followed in the last years which was focused on increasing the target mass to enhance the sensitivity to high WIMP masses.\\
Reducing the volume of the target crystal together with optimizing the detector layout allows to substantially gain in sensitivity for low WIMP masses with crystals of TUM40 quality. A schematic view of a possible detector design is presented in fig.~\ref{fig:Small_module}. The target crystal is a cuboid of dimensions (20$\times$20$\times$10)mm$^\text{3}$, corresponding to a target mass of 24g, and is paired with two light detectors of area (20$\times$20)mm$^2$, each facing a roughened face of the CaWO$_\text{4}$ crystal. The two light detectors and the CaWO$_\text{4}$ crystal are held by CaWO$_\text{4}$ sticks, giving a fully scintillating inner surface.
  
\begin{figure}[h!]
 \begin{center}
   \includegraphics[width=0.65\textwidth]{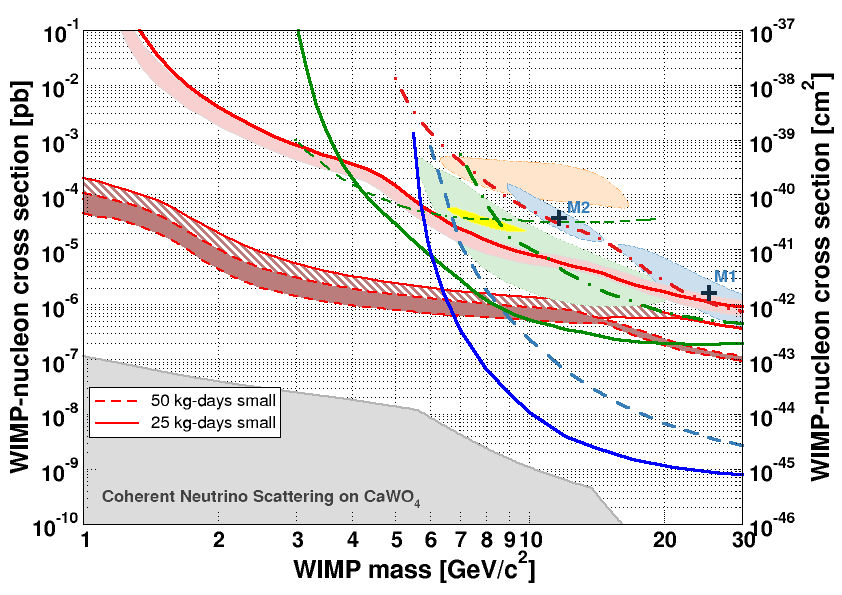}
   \caption{In addition to the limits shown in fig.~\ref{fig:Exclusion_run33}, the expected sensitivity (1 $\sigma$ C.L.) for different exposures of 24g detectors with the improved performance expected from the mass reduction of CaWO$_\text{4}$ of present quality.
   }
   \label{fig:Projections_small}
 \end{center}
\end{figure}

Given the size reduction of the target crystal and the presence of two light detectors covering a large solid angle, an increase in the detected light by a factor of 3 is plausible without changing the quality of the scintillating crystal. In addition we make the conservative assumption that the factor of 10 reduction in crystal volume will translate in a reduction of the threshold from the 0.6keV reached in \cite{CRESST_run33} to 0.1keV. Analogously, we assume that the reduction by a factor of 3 in the volume of the light detector absorber will result in a reduction of the light detector noise by a factor of 2.\\
With the listed premises and keeping the background and the non-proportionality of the crystals at the level of the crystal TUM40, we could reach a sensitivity that is shown in fig.~\ref{fig:Projections_small}. One year of measuring with 6 crystals of 24g each would allow us to reach an exposure of 50~kg~days where sensitivity will start being limited by detector performance.

\begin{figure}[h!]
 \begin{center}
   \includegraphics[width=0.65\textwidth]{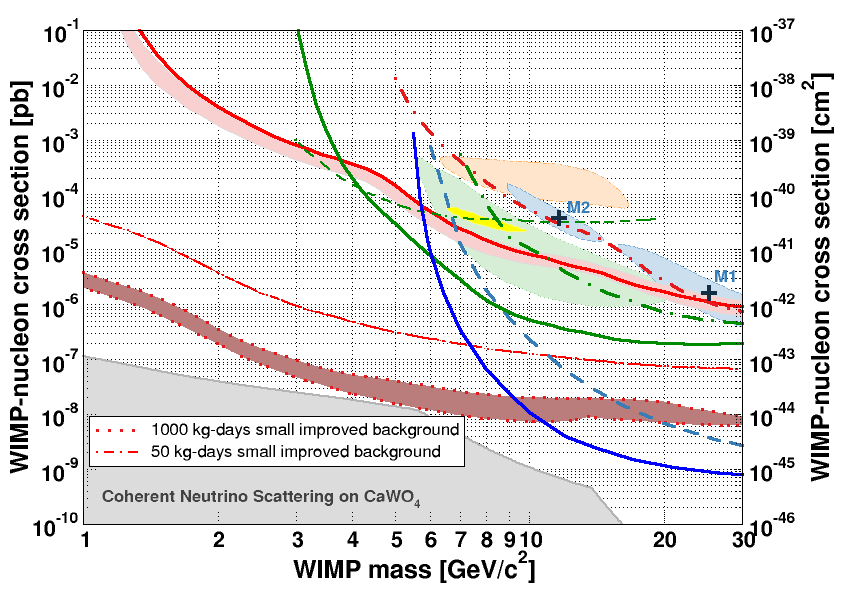}
   \caption{In addition to the limits shown in fig.~\ref{fig:Exclusion_run33}, the expected sensitivity (1 $\sigma$ C.L.) for different exposures of 24g detectors with the performances gain expected from size reduction and with $e^-\slash\gamma$ background reduced by a factor of 100.}
   \label{fig:Projections_small_final}
 \end{center}
\end{figure}

For any further improvement relative to the expectations shown in fig.~\ref{fig:Projections_small}, progress in crystal quality will be essential. This can be seen in fig.~\ref{fig:Projections_small_final} where, together with the assumptions for fig.~\ref{fig:Projections_small}, we considered a factor of 100 reduction in background. This goal will require an improved crystal radiopurity, a lower external radioactive contamination and a reduction of the background originated from the surrounding of the detectors.  With this additional gain in performance we could approach the coherent neutrino scattering limit with an exposure of 1000 kg days, which corresponds to about two years of data taking with 100 small modules (24g). We have to remark that, in order to accommodate such a 
moderate target mass in the existing CRESST facility, only an upgrade of the number of available read-out channels to about 300 would be needed.

\section{Next phases}
We are developing a phased program to approach the coherent neutrino scattering limit in 3 steps of increasing sensitivity. This strategy will allow us to proceed exploring the light WIMPs region with steps of up to 3 orders of magnitude in cross section sensitivity while we are accomplishing the activity needed for the following step.\\
The current run (run33) is planned to go on until enough statistics is achieved to completely explore the M1 region \cite{CRESST_run32}. In the mean time prototypes of small modules are produced and tested.\\
Key issues which need to be demonstrated in the next months before implementing the new modules in the main CRESST setup are: i) direct evaporation of TES on the absorber crystal (to avoid events from the carrier \cite{CRESST_run33}), ii) increased amount of detected light (possible reduction due to direct TES evaporation \cite{Michael_carrier}), iii) performance of the new holding system.\\

Once these decisive features of small modules are demonstrated, the upgrade to a new generation of the CRESST (CRESST III) experiment could proceed as follows:

\begin{itemize}
 
\item \textbf{Phase I}
 \begin{itemize}
    \item Physics Run:
      \begin{itemize}
            \item CRESST III-Run1 (run34) - run 10 small modules with crystals of available quality to reach the sensitivity estimated in fig.~\ref{fig:Projections_small};
      \end{itemize}
     \item R$\&$D for next phase:
       \begin{itemize}
            \item Revise design of standard-mass modules to improve discrimination of events from the carrier;
            \item Reach the desired background reduction and scintillation performance of the CaWO$_\text{4}$ crystals.
        \end{itemize}
 \end{itemize}
 Expected time scale: 06.2015-12.2016
 
\item \textbf{Phase II-a}
 \begin{itemize}
   \item Physics Run:
    \begin{itemize}
            \item CRESST III-Run2 (run35) - run again 10 small modules with improved crystals to finalize detector design and confirm crystal quality. This will allow us to reach the sensitivity shown in fig.~\ref{fig:Projections_small_final} in about one year of data taking;
     \end{itemize}
    \item R$\&$D for next phase:
     \begin{itemize}
      \item R\&D on large crystals for large-mass modules $O$(1kg) to be possibly used for gaining sensitivity at higher WIMP masses.
     \end{itemize}
\end{itemize}
 Expected time scale: 01.2017-12.2017
 
\item \textbf{Phase II-b}
 \begin{itemize}
  \item Full scale production of small detectors (100 modules of 24g);
  \item Upgrade of the cryostat:
  \begin{itemize}
   \item new detector support structure;
   \item 300 SQUIDs;
   \item new cabling for bias and heater lines;
   \item new electronics and DAQ.
  \end{itemize}
 \end{itemize}
 Expected time scale: 01.2018-12.2018
 
\item \textbf{Phase III}
\begin{itemize}
\item Physics Run:
 \begin{itemize}
  \item CRESST III-Run3 (run36) - run 100 small detector modules (2.4 kg of target mass) with improved crystal quality to reach, in two years, the sensitivity estimated in fig.~\ref{fig:Projections_small_final} and possibly a few large-mass modules to finalize the detector design (for a possible next step).
 \end{itemize}
\end{itemize}
Expected time scale: 01.2019-12.2020
\end{itemize}

\section{Summary}

This document marks a radical change of the previous strategy of the CRESST experiment. Recent experimental results have clearly shown that CRESST can play a leading role in exploring the low mass WIMP range in the field of dark matter searches. We present a well motivated and ambitious research and development program for the next years to further improve the sensitivity of the CRESST experiment. Current estimates based on realistic assumptions indicate an increase in the sensitivity for WIMP-nucleon elastic scattering cross section by about 2 to 3 orders of magnitude for a 1 to 6~GeV/c$^\text{2}$ WIMP and by more than 3 orders of magnitude for a 1~GeV/c$^\text{2}$ WIMP. Nevertheless, it has to be stressed that specializing for light WIMPs with masses lower than 10~GeV/c$^\text{2}$, does not exclude the possibility to use large-mass detectors to also gain sensitivity for higher WIMP masses (see fig.~\ref{fig:Projections_Run33}) and all developments will be performed in coordination with next 
generation dark matter experiments.

\section{Outlook}
The strategy presented here focuses on exploring the so-called low WIMP mass region, i.e. below 10~GeV/c$^\text{2}$. During the same time, liquid noble gas experiments will cover a large area of the high WIMP mass region above 10~GeV/c$^\text{2}$. In case of a discovery of WIMPs in the high mass region, the cryogenic technology exploited by CRESST will offer a mandatory and unique cross check of a possible signal.\\
For this reason it is crucial to keep the possibility to use the CRESST technology for a next generation dark matter experiment. In addition, a discovery of WIMPs in any mass range will open a new window for measurements, allowing a better understanding of the particle character of dark matter.\\
CRESST, as the only multi-element target experiment, is in an excellent position to significantly contribute to this next step of dark matter studies.\\
For this reason the CRESST collaboration will continue to study, in the framework of the EURECA/SuperCDMS collaboration, the possibility of an international large high resolution cryogenic facility to explore a wide range of WIMP masses with a multi-element target approach. All studies and detector developments performed during the next year will be integrated with ongoing studies in the context of the next generation of dark matter experiments.


\setlength{\bibsep}{2pt}

\end{document}